\documentclass[prl,twocolumn,superscriptaddress]{revtex4}
\usepackage{amsmath}
\usepackage{amssymb}
\usepackage{amsthm}
\usepackage{amsfonts}
\usepackage{listings}
\usepackage{enumerate}
\usepackage{latexsym}
\usepackage{color}
\usepackage{xcolor}
\usepackage{bm}
\usepackage{siunitx}
\usepackage{hyperref}
\hypersetup{
 pdfnewwindow=true, colorlinks=true,
 linkcolor=blue, anchorcolor=blue,
 citecolor=blue, filecolor=blue,
 menucolor=blue, urlcolor=blue}

\makeatletter

\newcommand{\Rmnum}[1]{\expandafter\@slowromancap\romannumeral #1@}
\makeatother

\usepackage{psfrag}

\usepackage{bm}
\usepackage{graphicx}
\usepackage{subfigure}

\usepackage{multirow}

\RequirePackage[normalem]{ulem} 
\RequirePackage{color}\definecolor{RED}{rgb}{1,0,0}\definecolor{BLUE}{rgb}{0,0,1}

\input{epsf}

\begin{document}
\title{Polarization-induced Quantum Spin Hall Insulator \\and Topological Devices in InAs Quantum Wells}

\author{Chenhao Liang}%
\affiliation{Beijing National Laboratory for Condensed Matter Physics and Institute of
Physics, Chinese Academy of Sciences, Beijing 100190, China}
\affiliation{University of Chinese Academy of Sciences, Beijing 100049, China}

\author{Sheng Zhang}
\affiliation{Beijing National Laboratory for Condensed Matter Physics and Institute of
Physics, Chinese Academy of Sciences, Beijing 100190, China}
\affiliation{University of Chinese Academy of Sciences, Beijing 100049, China}

\author{Haohao Sheng}
\affiliation{Beijing National Laboratory for Condensed Matter Physics and Institute of
Physics, Chinese Academy of Sciences, Beijing 100190, China}
\affiliation{University of Chinese Academy of Sciences, Beijing 100049, China}

\author{Quansheng~Wu}
\affiliation{Beijing National Laboratory for Condensed Matter Physics and Institute of
Physics, Chinese Academy of Sciences, Beijing 100190, China}
\affiliation{University of Chinese Academy of Sciences, Beijing 100049, China}

\author{Hongming Weng}
\affiliation{Beijing National Laboratory for Condensed Matter Physics and Institute of
Physics, Chinese Academy of Sciences, Beijing 100190, China}
\affiliation{University of Chinese Academy of Sciences, Beijing 100049, China}

\author{Zhong Fang}
\affiliation{Beijing National Laboratory for Condensed Matter Physics and Institute of
Physics, Chinese Academy of Sciences, Beijing 100190, China}
\affiliation{University of Chinese Academy of Sciences, Beijing 100049, China}

\author{Zhijun Wang}
\email{wzj@iphy.ac.cn}
\affiliation{Beijing National Laboratory for Condensed Matter Physics and Institute of
Physics, Chinese Academy of Sciences, Beijing 100190, China}
\affiliation{University of Chinese Academy of Sciences, Beijing 100049, China}

\date{\today}
\begin{abstract}
In this work, we predict the emergence of a quantum spin Hall insulator (QSHI) in conventional semiconductors, specifically InAs quantum wells, driven by a built-in polarization field. We propose QSHI InAs quantum wells as a platform to engineer topological field-effect devices. More precisely, we first present a novel topological logic device that operates without a topological phase transition. Subsequently, we design a high-performance topological transistor due to the presence of edge states. Our approach provides a potential framework for harnessing the unique features of QSHI in device design, paving the way for future topological devices.
\end{abstract}

\maketitle

\paragraph*{Introduction.}
Topological insulators are a novel class of materials distinguished by insulating bulk states and gapless conducting boundary states, which are protected by nontrivial topology~\cite{RevModPhys.88.021004}. 
The in-depth development of topological band theory has spurred extensive research into topological materials~\cite{bradlyn2017topological,po2017symmetry}, revealing that about 24\% of known materials exhibit topologically nontrivial phases~\cite{vergniory2019complete,tang2019comprehensive,zhang2019catalogue}.
In particular, a two-dimensional (2D) quantum spin Hall insulator (QSHI) ensures the formation of robust edge states that enable dissipationless transport, making it promising for applications~\cite{Narang2021,RevModPhys.83.1057}. It is of particular interest for next-generation electronic devices due to their compatibility with the trend towards device miniaturization and the enhanced electrostatic control inherent in low-dimensional systems~\cite{6606867,D3NR01288C,weber20242024}. 

Progress in topological devices~\cite{breunig2022opportunities} has been largely limited to topological field-effect transistors (FETs)~\cite{qian2014quantum,vandenberghe2017imperfect,nadeem2021overcoming} and spintronics~\cite{PhysRevLett.117.076802,huang2017bending,PhysRevLett.122.036401}, while topological logic devices are rare.
Usually, the opening of the gap is essential to control the transport of topological devices~\cite{vandenberghe2017imperfect}. Therefore, a topological phase transition or finite-size effect is necessary~\cite{10.1063/1.3268475,PhysRevLett.123.206801,PhysRevLett.101.246807,PhysRevB.88.121401}.
In order to improve performance at room temperature, topological devices rely on edge states, while bulk states need to be suppressed~\cite{breunig2022opportunities}. Consequently, a large bulk gap QSHI, such as 2D $X$enes ($X$=Bi, Sn and so on), is required in previous proprosals~\cite{molle2017buckled}. Furthermore, synthesizing high-quality samples of these materials poses challenges. To date, the electric-field-controlled topological phase transition has only been achieved in a few systems~\cite{PhysRevLett.130.196401,collinsElectricfieldtunedTopologicalPhase2018}.  How to effectively utilize the topological edge states of QSHI in device design remains an open question. 

In this letter, we find that QSHI can be achieved in InAs quantum wells (QWs) with a nontrivial gap of~$\sim$~50~meV. Its band inversion is driven by a built-in electric polarization field. An effective $k\cdot p$ Hamiltonian is obtained from \emph{ab-initio} calculations.  
Then, we propose two types of topological devices based on InAs QWs: a NOR logic gate and a high-performance FET.
In the NOR logic gate, which operates without a topological phase transition, the side gates are used to apply a transverse electric field to a narrow ribbon. 
In the FET device, the top split gate voltage is applied to regulate the effective width of the ribbon. Remarkably, its performance is comparable to that of state-of-the-art FETs.

\begin{figure}[!t]
	\centering
	\includegraphics[width=0.48\textwidth]{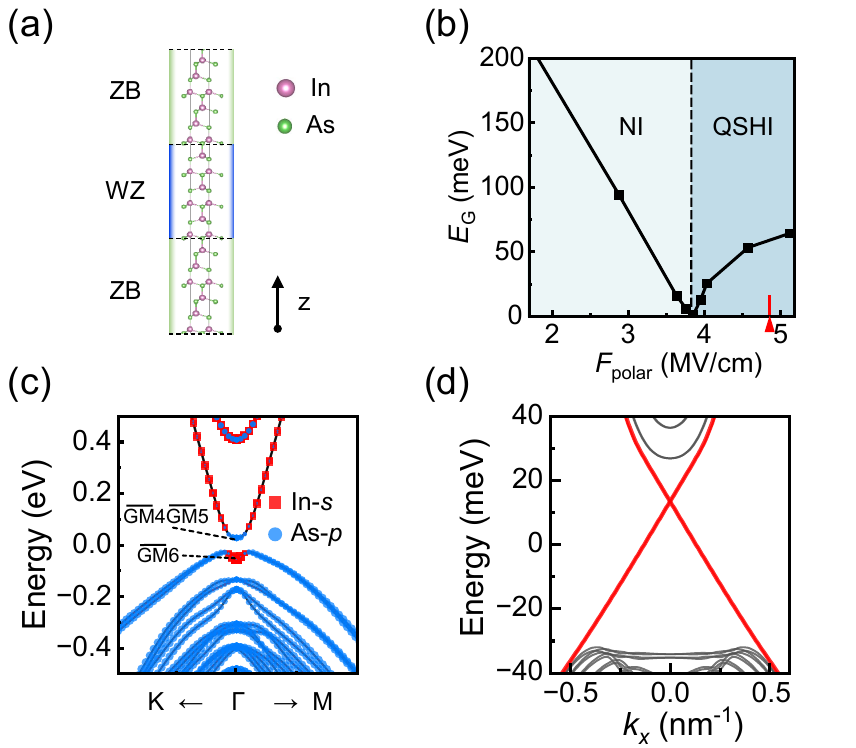}
	\caption{InAs QSHI and topological phase transition under polarization field.
    (a)~Structure of the InAs quantum wells.
    (b)~Bulk band gap $E_G$ and topological phase of InAs quantum wells as a function of polarization field $F_{\text{polar}}$ in WZ InAs. Both normal insulator (NI) and QSHI phase are shown. The red arrow marks the highest polarization field achieved in experiments~\cite{PhysRevB.97.115306}.
    (c)~Orbital-resolved DFT bulk band structures of a QSHI with polarization field $F_{\text{polar}}=4.58$ MV/cm.
    (d)~Band structure of this QSHI ribbon with width $W = 40$ nm from the $k\cdot p$ Hamiltonian. Red and black lines in (d) mark topological helical edge states and bulk states, respectively.
    }\label{fig-InAsTI}
\end{figure}

\paragraph*{Quantum spin Hall insulator in InAs quantum wells.}
In experiments, [0001]-wurtzite/[111]-zincblende (WZ/ZB) InAs quantum wells are successfully grown~\cite{10.1021/nl101632a,10.1021/nn504795v,nanolett.9b04507}. Due to the strain effect, a built-in polarization field is typically observed, reaching up to 4.90 MV/cm~\cite{dayehStructural2009,adma.201304021,PhysRevB.97.115306,PhysRevLett.109.186803,PhysRevLett.111.156402,PhysRevLett.112.216803}.
Here, we have systematically investigated the effect of the polarization field in WZ/ZB QWs. In our calculations, the QW structure consists of 2.1075 nm of WZ and 4.1932 nm of ZB InAs (in the $z$ direction) with the in-plane lattice constant $a=0.4279$ nm (in the $xy$ plane) as shown Fig.~\ref{fig-InAsTI}(a). Experiments demonstrate that the built-in polarization field is generated by the displacement of In atoms in the WZ structure along the [0001] direction, away from the center of the As$_4$ tetrahedron~\cite{adma.201304021}. For simplicity, we shift In atoms in the WZ phase of the QW structure to simulate the polarization field. 

Fig.~\ref{fig-InAsTI}(b) shows the topological phase transition of InAs QWs as the built-in polarization field~($F_{polar}$) varies in our first-principles calculations (see details in the Supplementary Materials (SM)~\cite{SupplementalMaterials}). The built-in $F_{\text{polar}}$ ($>3.85$ MV/cm) in QWs leads to band inversion and results in a QSHI. In the grown QW with $F_{\text{polar}}$=4.58 MV/cm~\cite{PhysRevB.97.115306,10.1021/nl101632a,adma.201304021}, it belongs to a QSHI with a 53 meV nontrivial band gap. The band inversion is clearly shown by the orbital-resolved DFT band structure in Fig.~\ref{fig-InAsTI}(c). The  $\mathbb{Z_\text{2}}=1$ index is calculated using the 1D Wilson loop method as implemented in the DFT package~\cite{Zhang_2023}. 

Furthermore, the irreducible representations of the low-energy bands are labeled as $\overline{\text{GM}}6$ (In $s$ orbital) and $\overline{\text{GM}}4\overline{\text{GM}}5$ (As $p$ orbitals) using IRVSP~\cite{GAO2021107760}, respectively.  Based on these, the low-energy effective Hamiltonian and the parameters are obtained by VASP2KP~\cite{Zhang_2023} (see the SM~\cite{SupplementalMaterials}). 
It is expressed as
 \begin{equation}
   \label{kp}
    H_{kp}=\epsilon(k)+ \begin{bmatrix}
        M(k)  & Ak_+         & -i\alpha k_-  & -i\beta k_+         \\
        Ak_- & -M(k)       & -i\beta k_+          & 0   \\
        i\alpha k_+  & i\beta k_-        & M(k)      & -Ak_- \\
        i\beta k_-          & 0        & -Ak_+      & -M(k)      \\
    \end{bmatrix} 
\end{equation}
where $\epsilon(k)=C+D(k_x^2+k_y^2)$, $M(k)=M+B(k_x^2+k_y^2)$, and $k_\pm=k_x\pm ik_y$. 
The parameters are $C=-0.020$~$\text{eV}$, $M=-0.04160$~$\text{eV}$, $D=0.3503$~$\text{eV}\cdot\text{nm}^2$, $B=0.4670$~$\text{eV}\cdot\text{nm}^2$, $A=-0.10835$~$\text{eV}\cdot\text{nm}$, $\alpha=-0.00484$~$\text{eV}\cdot\text{nm}$ and $\beta=0.11369$~$\text{eV}\cdot\text{nm}$. Accordingly, the topological helical edge states (red lines) are obtained in Fig.~\ref{fig-InAsTI}(d). 

\begin{figure}[!t]
	\centering
	\includegraphics[width=0.48\textwidth]{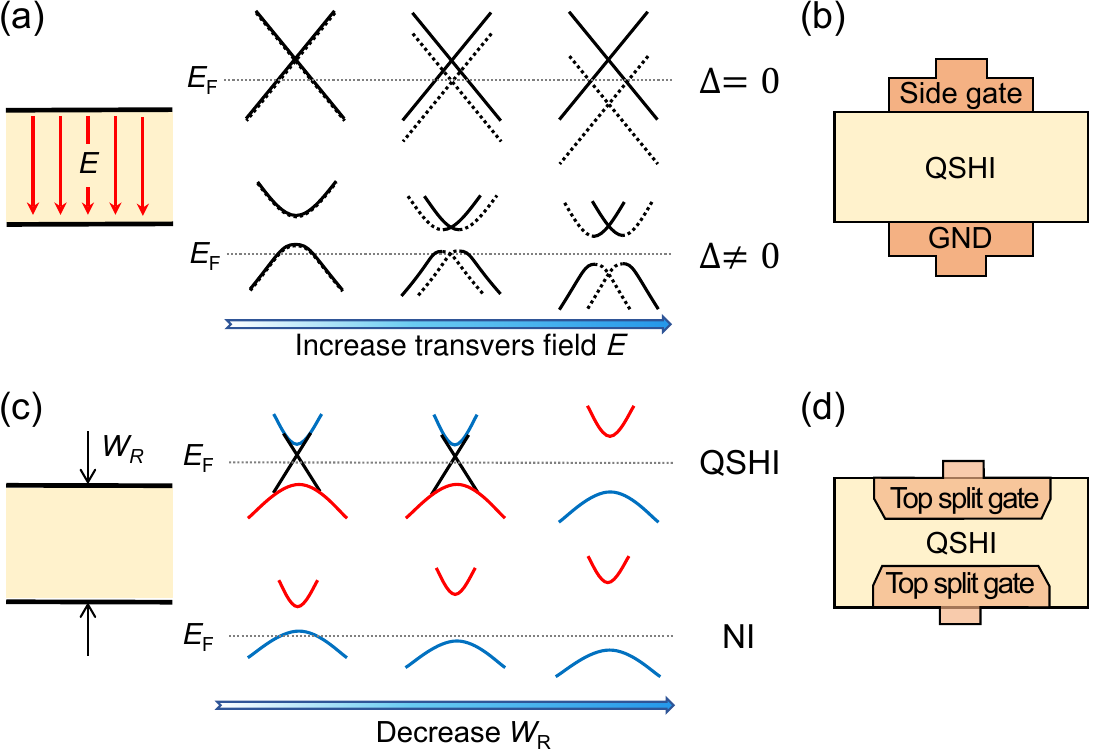}
	\caption{Gate control in topological insulator ribbon. (a)~A transverse electric field generated by the side gate $V_{\text{sg}}$ can shift the energy levels of the upper and lower edge states (dashed and solid lines). Thus, the energy level of the edge state gap can be tuned by a side gate in (b). (c)~The band structures of QSHI and NI exhibit completely different behaviors as the ribbon width narrows. The QSHI only exhibits a large gap in a narrow ribbon with strong coupling. The effective width $W_{R}$ can be reduced by top split gates in (d).}

    \label{fig-fieldonTI}
\end{figure}

\paragraph*{Design of topological devices in InAs QWs with electric field.}
The topological helical states emerge on the edges of the QSHI ribbon. When the ribbon becomes sufficiently narrow, the two edge states can couple with each other, and a hybridization gap opens. A transverse electric field, applied through side gates, can easily adjust the energy level of the gap. To illustrate the mechanism clearly, we construct an effective model for the coupled 1D edge states. It reads as 
\begin{equation}
   \label{1D-Hamiltonian}
H_{1D}(k_x)= \hbar v_F 
k_x\sigma_{z}\tau_{z}+\Delta \sigma_{0}\tau_{x}-eV \sigma_{0}\frac{(1+\tau_{z})}{2}
\end{equation}
where Pauli matrices $\sigma_{x,y,z}$ and $\tau_{x,y,z}$ represent the spin space and the edge space. $\sigma_0$ is the $2~\times~2$~unit matrix. $v_F$ is the Fermi velocity. $\hbar$ is the reduced Planck constant. The first term describes the topological helical edge states of the QSHI ribbon. The second term introduces the coupling strength $\Delta $ between the two edges due to the finite-size effect. The third term describes the electrostatic potential ($V$) on one edge (the other edge is set to zero).
For $\Delta=0$, although the energy levels of the edge states on one edge [dashed lines in Fig.~\ref{fig-fieldonTI}(a)] decrease as $V$ increases, the system remains metallic without any gap.
Once $\Delta\neq 0$, a hybridization gap is induced, and its energy level changes as $V$ varies, as shown in Fig.~\ref{fig-fieldonTI}(a). The system changes from a metallic state to an insulating state. This allows us to design a topological logic transistor with side gates of Fig.~\ref{fig-fieldonTI}(b) without a topological phase transition.

Additionally, due to the thermal excitation, topological devices based on a narrow gap QSHI perform poorly at room temperature. Usually, a large gap QSHI is required~\cite{breunig2022opportunities}. Previous studies have shown that top split gates can regulate the effective width of the ribbons~\cite{PhysRevLett.60.848,PhysRevB.83.081402,wuSpinpolarizedChargeTrapping2017a}. The evolutions of the band gap of the QSHI and NI are presented in Fig.~\ref{fig-fieldonTI}(c) as varying the ribbon width~($W_R$). In a 2D NI, there is always a band gap. However, a QSHI does not exhibit any gap in a relatively wide ribbon because of gapless edge states (black lines).  This can result in a high current in the $on$ state of a topological FET. The $on$ state exhibits dissipationless transport in the presence of nonmagnetic disorder, even at room temperature. This makes the topological FET of Fig.~\ref{fig-fieldonTI}(d) capable of achieving high performance at room temperature.

\begin{figure}[!t]
	\centering
	\includegraphics[width=0.48\textwidth]{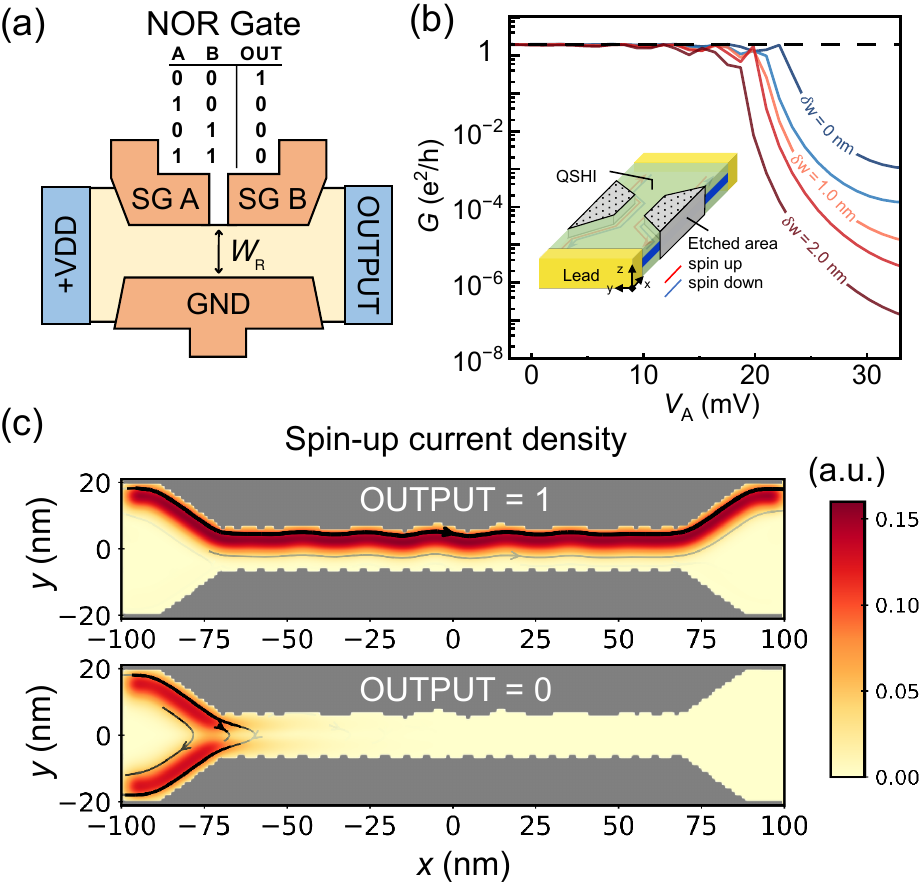}
	\caption{Topological logic transistor with the side gate. (a)~Schematic of the topological NOR gate and its truth table. (b)~Conductance as a function of voltage on side gate A $V_A$ in the presence of boundary-roughness $\delta$w. The black dashed line indicates $G=2e^2/h$. The inserted figure shows the device structure (without gates). 
 (c)~Spin-up current distribution at $\delta$w = 1.0 nm. $E_F$ is set to 0 meV. The width and length of the narrowed region are $W_{R}$ = 12 nm and $L_{R}$ = 140 nm. The length of side gate A or B is $L_{sg}$ = 65 nm.
    }\label{fig-NOR_gate}
\end{figure}

\paragraph*{Topological NOR logic gate.}
NOR/NAND logic gates are fundamental building blocks to perform complete logic operations in most logic systems~\cite{wakerly2008digital}, as all other logic operations can be derived from them.
Based on the transverse field effect of coupled edge states in the QSHI, we propose a NOR logic gate that operates without a topological phase transition, as illustrated in Fig.~\ref{fig-NOR_gate}(a). This logic gate utilizes coupled edge states with transverse electric fields applied through side gates (SG) A and B. In the state without voltage (logical inputs $V_A=0$ and $V_B=0$), the device is of a high-conductance $on$ state (logical output = 1). Most importantly, the output voltage is the same as the supply voltage $+VDD$ due to the dissipationless transport of topological edge states. Applying a voltage ($V_A=1$ or $V_B=1$) results in a low-conductance $off$ state (logical output = 0). Thus, it performs the complete NOR operation.

The device in Fig.~\ref{fig-NOR_gate}(a) has a width $W_{R}$ = 12 nm, to induce a hybridization gap ($\Delta_h=11.5$ meV). Since SG A and SG B are in series, we only consider $V_A$ and keep $V_B=0$ in the numerical calculations (see the SM~\cite{SupplementalMaterials}). As presented in Fig.~\ref{fig-NOR_gate}(b), when $V_A=0$, the calculated conductance is well-quantized at $2e^2/h$, which is consistent with the previous analysis that the topological edge states contribute to the quantized conductance.
When $V_A=30$ meV is applied, the conductance is reduced to $1.7\times 10^{-3}e^2/h$ in the $off$ state, as the Fermi level is in the hybridization gap. Most interestingly, once rough boundary~($\delta\text{w}$) is considered on the edges, the $on$ state is almost unchanged, as the topological edge states are robust against nonmagnetic impurities. However, in the $off$ state, the conductance is reduced to $2.7\times10^{-5}e^2/h$ at $\delta\text{w}=1.0$~nm,  and to $3.5\times10^{-7}e^2/h$ at $\delta\text{w}=2.0$~nm. In other words, the rough boundary in the device can enhance the switching performance. 

Fig.~\ref{fig-NOR_gate}(c) presents only the spin-up component of the spin-resolved current density and its direction in both the $on$ and $off$ states of our device with $\delta\text{w}=1.0$~nm.  It is clearly observed that in the $on$ state, the spin-polarized topological edge state current flows along the rough boundary and is localized at the upper edge without backscattering. This also clearly reveals the spin-moment locking nature of the topological edge states. In the $off$ state, due to the coupling of opposite edge states, carrier tunneling occurs from the upper edge to the lower edge, significantly reducing the conductance between the left and right terminals. Thus, the current is completely blocked. This transverse-field-controlled transport is unique to QSHIs, as time-reversal symmetry ensures that backscattering occurs only through tunneling between the two edges. Using a transverse electric field, one can control the topological edge current, enabling electric-field-controlled topological logic.

\begin{figure}[!t]
	\centering
	\includegraphics[width=0.48\textwidth]{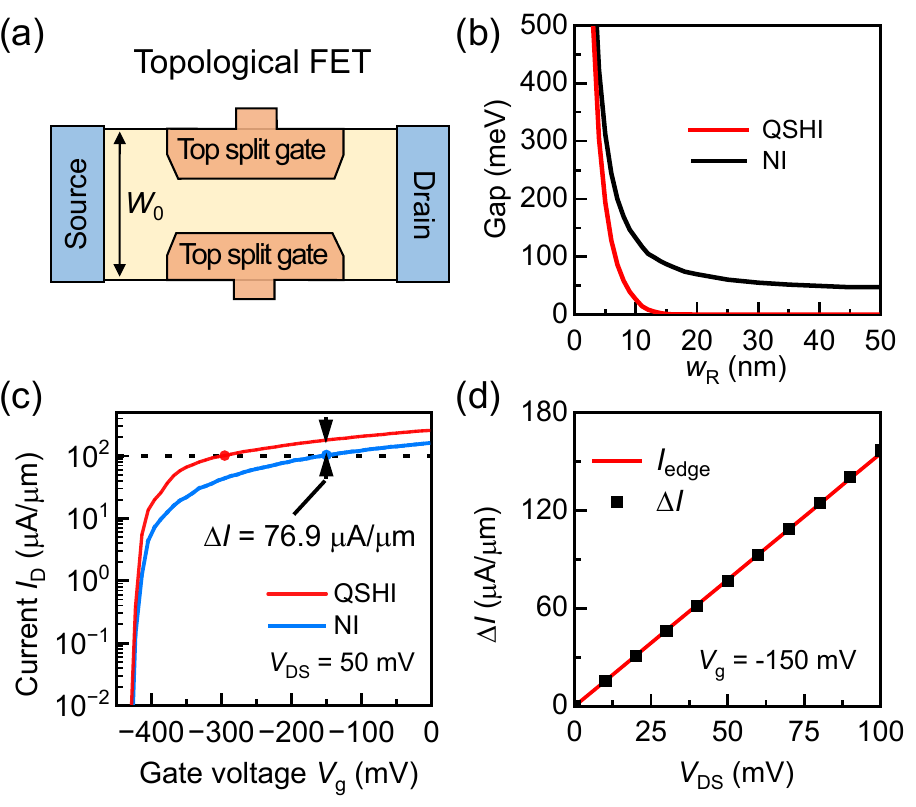}
	\caption{Topological FET with top split gates. (a) Schematic of the topological FET. The ribbon width is controlled by top gates. The width and length of a top gate are $W_{tg}=20$ nm and $L_{tg}$ = 140 nm. $W_0=50$ nm. (b) The evolution of band gaps for QSHI and NI with respect to ribbon width $W_{R}$.
    (c) Characteristic $I_{D}-V_{g}$ curve. The presence of edge states leads to better switching. Here, the the source-drain voltage $V_{DS}$ is set to 50 mV
    (d) The difference in current $\Delta I$ in (c) and the current $I_{D,\text{edge}}$ with quantum conductance $G_0=2e^2/h$ versus $V_{DS}$. 
    }\label{fig-low_ss}
\end{figure}

\paragraph*{Topological FET with top split gates.}
In addition to side gates, top split gates can also be used to regulate electron transport through the finite-size effect. We demonstrate that the existence of topological edge states in the width-tunable ribbon can significantly enhance the switching performance of a topological FET. In Fig.~\ref{fig-low_ss}(a), we design a topological FET with top split gates, where the effective~$W_{R}$ is controlled by the top gate voltage ($V_g$). Experimental studies have found that $W_{R}$ varies linearly with $V_g$ over a wide range of 0 to 200 nm~\cite{PhysRevLett.60.848}. Thus, it can be parameterized as $W_{R}=W_{0} + \alpha V_{g}$, with the ribbon width $W_0=50~\text{nm}$ and $\alpha=110~\text{nm/V}$ in~Fig.~\ref{fig-low_ss}.
Fig.~\ref{fig-low_ss}(b) illustrates the band gaps for the QSHI and the NI~\cite{NI_para} with different $W_{R}$. In the NI ribbon, there is always an energy gap. However, it starts to open an observable edge state gap ($\sim~1.2~\text{meV}$) in the QSHI ribbon, when $W_{R}$ decreases to 15 nm. Once the $W_{R}$ becomes less than 15 nm, the gap increases rapidly.

In Fig.~\ref{fig-low_ss}(c), we simulate the characteristic $I_D - V_{g}$ curves of the topological FET at room temperature ($T$ = 300 K) and the source-drain voltage $V_{\text{DS}}$ = 50 mV.
The $I_D - V_{g}$ curve of a NI FET with the same device structure is marked by the blue line in Fig.~\ref{fig-low_ss}(c) for comparison. The critical $on$- (resp. $off$-) state current of the FET is $I_{on} = 10^2$~$\ \mu\text{A}/\mu\text{m}$  (resp. $I_{off} = 10^{-2}$~$\ \mu\text{A}/\mu\text{m}$). We can see that at the same gate voltage $V_{\text{g}}$, the $I_D$ of the topological FET is higher. The gate voltage change required to switch the topological FET is approximately reduced by 50\% compared to that of the NI FET. To quantitatively assess performance improvement, we introduce a commonly used performance metric in FETs: subthreshold swing $SS=\partial V_{g}/\partial \log{I_D}$~\cite{datta}. For conventional FETs at room temperature, the theoretical minimum is $SS_{min}=60$~mV/decade~\cite{kim2020thickness}. The average subthreshold swing $SS_{\text{ave(4)}}$ (current changes over $10^4$), in our devices, decreases from 69.5 mV/dec to 33.5 mV/dec, breaking the theoretical limit of subthreshold swing. This $SS_{\text{ave(4)}}$ is comparable to advanced tunneling FET, negative-capacity FET~\cite{kamaei2023ferroelectric} and Dirac-source FET~\cite{acsnano.1c01503}. 

It is worthwhile to notice the current difference $\Delta I$ between the topological FET and the NI FET. Fig.~\ref{fig-low_ss}(c) shows that $\Delta I$ is 76.9 $\mu\text{A}/\mu\text{m}$ at $V_{g} = -150$ mV. It corresponds to the topological edge current. We can compare \(\Delta I\) with the current contributed solely by the topological edge state, \(I_{\text{edge}}=G_0 V_{DS}\), where $G_0=2e^2/h$. In Fig.~\ref{fig-low_ss}(d), the numerical result of $\Delta I$ at different $V_{\text{DS}}$ is shown. \(\Delta I\) is close to $I_{\text{edge}}$ at $V_{\text{DS}}$ up to 100 mV. Although the bulk gap of this narrow gap QSHI is only about 50 meV, the contribution of the edge states can still be visualized at $T=300$ K. With nonmagnetic impurities, we conclude that the bulk current may be suppressed while the topological edge states in the $on$ state are preserved. Thus, topological FETs can maintain promising switching characteristics with nonmagnetic impurities. This suggests that a large bulk gap (e.g., over 100 meV) QSHI may not be necessary to achieve high-performance topological devices.

\paragraph*{Discussion.}

In this work, we find that the band structure can be tuned by the built-in electric polarization field of grown InAs QWs. Accordingly, a narrow-gap QSHI can be achieved with the experimental $F_{polar}$.  We investigate the transport properties of the QSHI InAs ribbon with a field-controlled finite-size effect. Using the topological edge states of QSHI InAs QWs, we propose a topological logic without a topological transition and a high-performance topological FET even at room temperature. As device miniaturization continues, research on the transport properties in quasi-one-dimensional channels is an important topic for topological devices. Our finding provides new insight into the designs of the future topological devices. 

The built-in polarization in the grown samples is widely found in narrow-gap III-V semiconductors.
Besides the WZ/ZB InAs QWs~\cite{10.1021/nl101632a,10.1021/nn504795v}, high-quality WZ/ZB InSb and built-in polarization are also obtained experimentally~\cite{ANANDAN201930}. Selective area epitaxy has enabled the fabrication of shape-controllable WZ/ZB InAs heterostructures~\cite{nanolett.9b04507} and quantum well arrays~\cite{10.1021/acsnano.1c00483}. This makes these topological devices achievable. In addition, the previous examples of topological logic have recently been realized in acoustic and photonic systems~\cite{pirie_topological_2022,zhang_experimental_2023}. The former relies on temperature and stress to control the topological phase, while the latter uses the phase difference of light. Our approach simplifies device design and integration by avoiding the complex conversion of physical quantities. Moreover, all inputs and outputs can be defined by the same supplied voltage in this topological NOR gate.  This consistency in physical quantities facilitates the cascading of devices.

\paragraph*{Acknowledgements.}
This work was supported by the National Natural Science Foundation of China (Grants No. 12188101), National Key R\&D Program of China (Grants No. 2022YFA1403800), and the Center for Materials Genome.


\begin{thebibliography}{56}
\expandafter\ifx\csname natexlab\endcsname\relax\def\natexlab#1{#1}\fi
\expandafter\ifx\csname bibnamefont\endcsname\relax
  \def\bibnamefont#1{#1}\fi
\expandafter\ifx\csname bibfnamefont\endcsname\relax
  \def\bibfnamefont#1{#1}\fi
\expandafter\ifx\csname citenamefont\endcsname\relax
  \def\citenamefont#1{#1}\fi
\expandafter\ifx\csname url\endcsname\relax
  \def\url#1{\texttt{#1}}\fi
\expandafter\ifx\csname urlprefix\endcsname\relax\def\urlprefix{URL }\fi
\providecommand{\bibinfo}[2]{#2}
\providecommand{\eprint}[2][]{\url{#2}}

\bibitem[{\citenamefont{Bansil et~al.}(2016)\citenamefont{Bansil, Lin, and Das}}]{RevModPhys.88.021004}
\bibinfo{author}{\bibfnamefont{A.}~\bibnamefont{Bansil}}, \bibinfo{author}{\bibfnamefont{H.}~\bibnamefont{Lin}}, \bibnamefont{and} \bibinfo{author}{\bibfnamefont{T.}~\bibnamefont{Das}}, \bibinfo{journal}{Rev. Mod. Phys.} \textbf{\bibinfo{volume}{88}}, \bibinfo{pages}{021004} (\bibinfo{year}{2016}), \urlprefix\url{https://link.aps.org/doi/10.1103/RevModPhys.88.021004}.

\bibitem[{\citenamefont{Bradlyn et~al.}(2017)\citenamefont{Bradlyn, Elcoro, Cano, Vergniory, Wang, Felser, Aroyo, and Bernevig}}]{bradlyn2017topological}
\bibinfo{author}{\bibfnamefont{B.}~\bibnamefont{Bradlyn}}, \bibinfo{author}{\bibfnamefont{L.}~\bibnamefont{Elcoro}}, \bibinfo{author}{\bibfnamefont{J.}~\bibnamefont{Cano}}, \bibinfo{author}{\bibfnamefont{M.~G.} \bibnamefont{Vergniory}}, \bibinfo{author}{\bibfnamefont{Z.}~\bibnamefont{Wang}}, \bibinfo{author}{\bibfnamefont{C.}~\bibnamefont{Felser}}, \bibinfo{author}{\bibfnamefont{M.~I.} \bibnamefont{Aroyo}}, \bibnamefont{and} \bibinfo{author}{\bibfnamefont{B.~A.} \bibnamefont{Bernevig}}, \bibinfo{journal}{Nature} \textbf{\bibinfo{volume}{547}}, \bibinfo{pages}{298} (\bibinfo{year}{2017}), \urlprefix\url{https://doi.org/10.1038/nature23268}.

\bibitem[{\citenamefont{Po et~al.}(2017)\citenamefont{Po, Vishwanath, and Watanabe}}]{po2017symmetry}
\bibinfo{author}{\bibfnamefont{H.~C.} \bibnamefont{Po}}, \bibinfo{author}{\bibfnamefont{A.}~\bibnamefont{Vishwanath}}, \bibnamefont{and} \bibinfo{author}{\bibfnamefont{H.}~\bibnamefont{Watanabe}}, \bibinfo{journal}{Nature communications} \textbf{\bibinfo{volume}{8}}, \bibinfo{pages}{50} (\bibinfo{year}{2017}), \urlprefix\url{https://doi.org/10.1038/s41467-017-00133-2}.

\bibitem[{\citenamefont{Vergniory et~al.}(2019)\citenamefont{Vergniory, Elcoro, Felser, Regnault, Bernevig, and Wang}}]{vergniory2019complete}
\bibinfo{author}{\bibfnamefont{M.}~\bibnamefont{Vergniory}}, \bibinfo{author}{\bibfnamefont{L.}~\bibnamefont{Elcoro}}, \bibinfo{author}{\bibfnamefont{C.}~\bibnamefont{Felser}}, \bibinfo{author}{\bibfnamefont{N.}~\bibnamefont{Regnault}}, \bibinfo{author}{\bibfnamefont{B.~A.} \bibnamefont{Bernevig}}, \bibnamefont{and} \bibinfo{author}{\bibfnamefont{Z.}~\bibnamefont{Wang}}, \bibinfo{journal}{Nature} \textbf{\bibinfo{volume}{566}}, \bibinfo{pages}{480} (\bibinfo{year}{2019}), \urlprefix\url{https://doi.org/10.1038/s41586-019-0954-4}.

\bibitem[{\citenamefont{Tang et~al.}(2019)\citenamefont{Tang, Po, Vishwanath, and Wan}}]{tang2019comprehensive}
\bibinfo{author}{\bibfnamefont{F.}~\bibnamefont{Tang}}, \bibinfo{author}{\bibfnamefont{H.~C.} \bibnamefont{Po}}, \bibinfo{author}{\bibfnamefont{A.}~\bibnamefont{Vishwanath}}, \bibnamefont{and} \bibinfo{author}{\bibfnamefont{X.}~\bibnamefont{Wan}}, \bibinfo{journal}{Nature} \textbf{\bibinfo{volume}{566}}, \bibinfo{pages}{486} (\bibinfo{year}{2019}), \urlprefix\url{https://doi.org/10.1038/s41586-019-0937-5}.

\bibitem[{\citenamefont{Zhang et~al.}(2019)\citenamefont{Zhang, Jiang, Song, Huang, He, Fang, Weng, and Fang}}]{zhang2019catalogue}
\bibinfo{author}{\bibfnamefont{T.}~\bibnamefont{Zhang}}, \bibinfo{author}{\bibfnamefont{Y.}~\bibnamefont{Jiang}}, \bibinfo{author}{\bibfnamefont{Z.}~\bibnamefont{Song}}, \bibinfo{author}{\bibfnamefont{H.}~\bibnamefont{Huang}}, \bibinfo{author}{\bibfnamefont{Y.}~\bibnamefont{He}}, \bibinfo{author}{\bibfnamefont{Z.}~\bibnamefont{Fang}}, \bibinfo{author}{\bibfnamefont{H.}~\bibnamefont{Weng}}, \bibnamefont{and} \bibinfo{author}{\bibfnamefont{C.}~\bibnamefont{Fang}}, \bibinfo{journal}{Nature} \textbf{\bibinfo{volume}{566}}, \bibinfo{pages}{475} (\bibinfo{year}{2019}), \urlprefix\url{https://doi.org/10.1038/s41586-019-0944-6}.

\bibitem[{\citenamefont{Narang et~al.}(2021)\citenamefont{Narang, Garcia, and Felser}}]{Narang2021}
\bibinfo{author}{\bibfnamefont{P.}~\bibnamefont{Narang}}, \bibinfo{author}{\bibfnamefont{C.~A.} \bibnamefont{Garcia}}, \bibnamefont{and} \bibinfo{author}{\bibfnamefont{C.}~\bibnamefont{Felser}}, \bibinfo{journal}{Nature Materials} \textbf{\bibinfo{volume}{20}}, \bibinfo{pages}{293} (\bibinfo{year}{2021}), \urlprefix\url{https://doi.org/10.1038/s41563-020-00820-4}.

\bibitem[{\citenamefont{Qi and Zhang}(2011)}]{RevModPhys.83.1057}
\bibinfo{author}{\bibfnamefont{X.-L.} \bibnamefont{Qi}} \bibnamefont{and} \bibinfo{author}{\bibfnamefont{S.-C.} \bibnamefont{Zhang}}, \bibinfo{journal}{Rev. Mod. Phys.} \textbf{\bibinfo{volume}{83}}, \bibinfo{pages}{1057} (\bibinfo{year}{2011}), \urlprefix\url{https://link.aps.org/doi/10.1103/RevModPhys.83.1057}.

\bibitem[{\citenamefont{Fischetti et~al.}(2013)\citenamefont{Fischetti, Fu, and Vandenberghe}}]{6606867}
\bibinfo{author}{\bibfnamefont{M.~V.} \bibnamefont{Fischetti}}, \bibinfo{author}{\bibfnamefont{B.}~\bibnamefont{Fu}}, \bibnamefont{and} \bibinfo{author}{\bibfnamefont{W.~G.} \bibnamefont{Vandenberghe}}, \bibinfo{journal}{IEEE Transactions on Electron Devices} \textbf{\bibinfo{volume}{60}}, \bibinfo{pages}{3862} (\bibinfo{year}{2013}), \urlprefix\url{https://doi.org/10.1109/TED.2013.2280844}.

\bibitem[{\citenamefont{Jin et~al.}(2023)\citenamefont{Jin, Jiang, Sethi, and Liu}}]{D3NR01288C}
\bibinfo{author}{\bibfnamefont{K.-H.} \bibnamefont{Jin}}, \bibinfo{author}{\bibfnamefont{W.}~\bibnamefont{Jiang}}, \bibinfo{author}{\bibfnamefont{G.}~\bibnamefont{Sethi}}, \bibnamefont{and} \bibinfo{author}{\bibfnamefont{F.}~\bibnamefont{Liu}}, \bibinfo{journal}{Nanoscale} \textbf{\bibinfo{volume}{15}}, \bibinfo{pages}{12787} (\bibinfo{year}{2023}), \urlprefix\url{http://dx.doi.org/10.1039/D3NR01288C}.

\bibitem[{\citenamefont{Weber et~al.}(2024)\citenamefont{Weber, Fuhrer, Sheng, Yang, Thomale, Shamim, Molenkamp, Cobden, Pesin, Zandvliet et~al.}}]{weber20242024}
\bibinfo{author}{\bibfnamefont{B.}~\bibnamefont{Weber}}, \bibinfo{author}{\bibfnamefont{M.}~\bibnamefont{Fuhrer}}, \bibinfo{author}{\bibfnamefont{X.-L.} \bibnamefont{Sheng}}, \bibinfo{author}{\bibfnamefont{S.~A.} \bibnamefont{Yang}}, \bibinfo{author}{\bibfnamefont{R.}~\bibnamefont{Thomale}}, \bibinfo{author}{\bibfnamefont{S.}~\bibnamefont{Shamim}}, \bibinfo{author}{\bibfnamefont{L.~W.} \bibnamefont{Molenkamp}}, \bibinfo{author}{\bibfnamefont{D.~H.} \bibnamefont{Cobden}}, \bibinfo{author}{\bibfnamefont{D.}~\bibnamefont{Pesin}}, \bibinfo{author}{\bibfnamefont{H.~J.} \bibnamefont{Zandvliet}}, \bibnamefont{et~al.}, \bibinfo{journal}{Journal of Physics: Materials}  (\bibinfo{year}{2024}), \urlprefix\url{https://doi.org/10.1088/2515-7639/ad2083}.

\bibitem[{\citenamefont{Breunig and Ando}(2022)}]{breunig2022opportunities}
\bibinfo{author}{\bibfnamefont{O.}~\bibnamefont{Breunig}} \bibnamefont{and} \bibinfo{author}{\bibfnamefont{Y.}~\bibnamefont{Ando}}, \bibinfo{journal}{Nature Reviews Physics} \textbf{\bibinfo{volume}{4}}, \bibinfo{pages}{184} (\bibinfo{year}{2022}), \urlprefix\url{https://doi.org/10.1038/s42254-021-00402-6}.

\bibitem[{\citenamefont{Qian et~al.}(2014)\citenamefont{Qian, Liu, Fu, and Li}}]{qian2014quantum}
\bibinfo{author}{\bibfnamefont{X.}~\bibnamefont{Qian}}, \bibinfo{author}{\bibfnamefont{J.}~\bibnamefont{Liu}}, \bibinfo{author}{\bibfnamefont{L.}~\bibnamefont{Fu}}, \bibnamefont{and} \bibinfo{author}{\bibfnamefont{J.}~\bibnamefont{Li}}, \bibinfo{journal}{Science} \textbf{\bibinfo{volume}{346}}, \bibinfo{pages}{1344} (\bibinfo{year}{2014}), \urlprefix\url{https://www.science.org/doi/full/10.1126/science.1256815}.

\bibitem[{\citenamefont{Vandenberghe and Fischetti}(2017)}]{vandenberghe2017imperfect}
\bibinfo{author}{\bibfnamefont{W.~G.} \bibnamefont{Vandenberghe}} \bibnamefont{and} \bibinfo{author}{\bibfnamefont{M.~V.} \bibnamefont{Fischetti}}, \bibinfo{journal}{Nature communications} \textbf{\bibinfo{volume}{8}}, \bibinfo{pages}{14184} (\bibinfo{year}{2017}), \urlprefix\url{https://doi.org/10.1038/ncomms14184}.

\bibitem[{\citenamefont{Nadeem et~al.}(2021)\citenamefont{Nadeem, Di~Bernardo, Wang, Fuhrer, and Culcer}}]{nadeem2021overcoming}
\bibinfo{author}{\bibfnamefont{M.}~\bibnamefont{Nadeem}}, \bibinfo{author}{\bibfnamefont{I.}~\bibnamefont{Di~Bernardo}}, \bibinfo{author}{\bibfnamefont{X.}~\bibnamefont{Wang}}, \bibinfo{author}{\bibfnamefont{M.~S.} \bibnamefont{Fuhrer}}, \bibnamefont{and} \bibinfo{author}{\bibfnamefont{D.}~\bibnamefont{Culcer}}, \bibinfo{journal}{Nano Letters} \textbf{\bibinfo{volume}{21}}, \bibinfo{pages}{3155} (\bibinfo{year}{2021}), \urlprefix\url{https://pubs.acs.org/doi/full/10.1021/acs.nanolett.1c00378}.

\bibitem[{\citenamefont{Chen et~al.}(2016)\citenamefont{Chen, Deng, Hou, Shi, Sheng, and Xing}}]{PhysRevLett.117.076802}
\bibinfo{author}{\bibfnamefont{W.}~\bibnamefont{Chen}}, \bibinfo{author}{\bibfnamefont{W.-Y.} \bibnamefont{Deng}}, \bibinfo{author}{\bibfnamefont{J.-M.} \bibnamefont{Hou}}, \bibinfo{author}{\bibfnamefont{D.~N.} \bibnamefont{Shi}}, \bibinfo{author}{\bibfnamefont{L.}~\bibnamefont{Sheng}}, \bibnamefont{and} \bibinfo{author}{\bibfnamefont{D.~Y.} \bibnamefont{Xing}}, \bibinfo{journal}{Phys. Rev. Lett.} \textbf{\bibinfo{volume}{117}}, \bibinfo{pages}{076802} (\bibinfo{year}{2016}), \urlprefix\url{https://link.aps.org/doi/10.1103/PhysRevLett.117.076802}.

\bibitem[{\citenamefont{Huang et~al.}(2017)\citenamefont{Huang, Jin, Cui, Zhai, Mei, and Liu}}]{huang2017bending}
\bibinfo{author}{\bibfnamefont{B.}~\bibnamefont{Huang}}, \bibinfo{author}{\bibfnamefont{K.-H.} \bibnamefont{Jin}}, \bibinfo{author}{\bibfnamefont{B.}~\bibnamefont{Cui}}, \bibinfo{author}{\bibfnamefont{F.}~\bibnamefont{Zhai}}, \bibinfo{author}{\bibfnamefont{J.}~\bibnamefont{Mei}}, \bibnamefont{and} \bibinfo{author}{\bibfnamefont{F.}~\bibnamefont{Liu}}, \bibinfo{journal}{Nature Communications} \textbf{\bibinfo{volume}{8}}, \bibinfo{pages}{1} (\bibinfo{year}{2017}), \urlprefix\url{https://doi.org/10.1038/ncomms15850}.

\bibitem[{\citenamefont{Acosta and Fazzio}(2019)}]{PhysRevLett.122.036401}
\bibinfo{author}{\bibfnamefont{C.~M.} \bibnamefont{Acosta}} \bibnamefont{and} \bibinfo{author}{\bibfnamefont{A.}~\bibnamefont{Fazzio}}, \bibinfo{journal}{Phys. Rev. Lett.} \textbf{\bibinfo{volume}{122}}, \bibinfo{pages}{036401} (\bibinfo{year}{2019}), \urlprefix\url{https://link.aps.org/doi/10.1103/PhysRevLett.122.036401}.

\bibitem[{\citenamefont{Li and Chang}(2009)}]{10.1063/1.3268475}
\bibinfo{author}{\bibfnamefont{J.}~\bibnamefont{Li}} \bibnamefont{and} \bibinfo{author}{\bibfnamefont{K.}~\bibnamefont{Chang}}, \bibinfo{journal}{Applied Physics Letters} \textbf{\bibinfo{volume}{95}}, \bibinfo{pages}{222110} (\bibinfo{year}{2009}), \urlprefix\url{https://doi.org/10.1063/1.3268475}.

\bibitem[{\citenamefont{Xu et~al.}(2019)\citenamefont{Xu, Chen, Wang, Liu, and Ma}}]{PhysRevLett.123.206801}
\bibinfo{author}{\bibfnamefont{Y.}~\bibnamefont{Xu}}, \bibinfo{author}{\bibfnamefont{Y.-R.} \bibnamefont{Chen}}, \bibinfo{author}{\bibfnamefont{J.}~\bibnamefont{Wang}}, \bibinfo{author}{\bibfnamefont{J.-F.} \bibnamefont{Liu}}, \bibnamefont{and} \bibinfo{author}{\bibfnamefont{Z.}~\bibnamefont{Ma}}, \bibinfo{journal}{Phys. Rev. Lett.} \textbf{\bibinfo{volume}{123}}, \bibinfo{pages}{206801} (\bibinfo{year}{2019}), \urlprefix\url{https://link.aps.org/doi/10.1103/PhysRevLett.123.206801}.

\bibitem[{\citenamefont{Zhou et~al.}(2008)\citenamefont{Zhou, Lu, Chu, Shen, and Niu}}]{PhysRevLett.101.246807}
\bibinfo{author}{\bibfnamefont{B.}~\bibnamefont{Zhou}}, \bibinfo{author}{\bibfnamefont{H.-Z.} \bibnamefont{Lu}}, \bibinfo{author}{\bibfnamefont{R.-L.} \bibnamefont{Chu}}, \bibinfo{author}{\bibfnamefont{S.-Q.} \bibnamefont{Shen}}, \bibnamefont{and} \bibinfo{author}{\bibfnamefont{Q.}~\bibnamefont{Niu}}, \bibinfo{journal}{Phys. Rev. Lett.} \textbf{\bibinfo{volume}{101}}, \bibinfo{pages}{246807} (\bibinfo{year}{2008}), \urlprefix\url{https://link.aps.org/doi/10.1103/PhysRevLett.101.246807}.

\bibitem[{\citenamefont{Ezawa and Nagaosa}(2013)}]{PhysRevB.88.121401}
\bibinfo{author}{\bibfnamefont{M.}~\bibnamefont{Ezawa}} \bibnamefont{and} \bibinfo{author}{\bibfnamefont{N.}~\bibnamefont{Nagaosa}}, \bibinfo{journal}{Phys. Rev. B} \textbf{\bibinfo{volume}{88}}, \bibinfo{pages}{121401} (\bibinfo{year}{2013}), \urlprefix\url{https://link.aps.org/doi/10.1103/PhysRevB.88.121401}.

\bibitem[{\citenamefont{Molle et~al.}(2017)\citenamefont{Molle, Goldberger, Houssa, Xu, Zhang, and Akinwande}}]{molle2017buckled}
\bibinfo{author}{\bibfnamefont{A.}~\bibnamefont{Molle}}, \bibinfo{author}{\bibfnamefont{J.}~\bibnamefont{Goldberger}}, \bibinfo{author}{\bibfnamefont{M.}~\bibnamefont{Houssa}}, \bibinfo{author}{\bibfnamefont{Y.}~\bibnamefont{Xu}}, \bibinfo{author}{\bibfnamefont{S.-C.} \bibnamefont{Zhang}}, \bibnamefont{and} \bibinfo{author}{\bibfnamefont{D.}~\bibnamefont{Akinwande}}, \bibinfo{journal}{Nature materials} \textbf{\bibinfo{volume}{16}}, \bibinfo{pages}{163} (\bibinfo{year}{2017}), \urlprefix\url{https://doi.org/10.1038/nmat4802}.

\bibitem[{\citenamefont{Bampoulis et~al.}(2023)\citenamefont{Bampoulis, Castenmiller, Klaassen, van Mil, Liu, Liu, Yao, Ezawa, Rudenko, and Zandvliet}}]{PhysRevLett.130.196401}
\bibinfo{author}{\bibfnamefont{P.}~\bibnamefont{Bampoulis}}, \bibinfo{author}{\bibfnamefont{C.}~\bibnamefont{Castenmiller}}, \bibinfo{author}{\bibfnamefont{D.~J.} \bibnamefont{Klaassen}}, \bibinfo{author}{\bibfnamefont{J.}~\bibnamefont{van Mil}}, \bibinfo{author}{\bibfnamefont{Y.}~\bibnamefont{Liu}}, \bibinfo{author}{\bibfnamefont{C.-C.} \bibnamefont{Liu}}, \bibinfo{author}{\bibfnamefont{Y.}~\bibnamefont{Yao}}, \bibinfo{author}{\bibfnamefont{M.}~\bibnamefont{Ezawa}}, \bibinfo{author}{\bibfnamefont{A.~N.} \bibnamefont{Rudenko}}, \bibnamefont{and} \bibinfo{author}{\bibfnamefont{H.~J.~W.} \bibnamefont{Zandvliet}}, \bibinfo{journal}{Phys. Rev. Lett.} \textbf{\bibinfo{volume}{130}}, \bibinfo{pages}{196401} (\bibinfo{year}{2023}), \urlprefix\url{https://link.aps.org/doi/10.1103/PhysRevLett.130.196401}.

\bibitem[{\citenamefont{Collins et~al.}(2018)\citenamefont{Collins, Tadich, Wu, Gomes, Rodrigues, Liu, Hellerstedt, Ryu, Tang, Mo et~al.}}]{collinsElectricfieldtunedTopologicalPhase2018}
\bibinfo{author}{\bibfnamefont{J.~L.} \bibnamefont{Collins}}, \bibinfo{author}{\bibfnamefont{A.}~\bibnamefont{Tadich}}, \bibinfo{author}{\bibfnamefont{W.}~\bibnamefont{Wu}}, \bibinfo{author}{\bibfnamefont{L.~C.} \bibnamefont{Gomes}}, \bibinfo{author}{\bibfnamefont{J.~N.~B.} \bibnamefont{Rodrigues}}, \bibinfo{author}{\bibfnamefont{C.}~\bibnamefont{Liu}}, \bibinfo{author}{\bibfnamefont{J.}~\bibnamefont{Hellerstedt}}, \bibinfo{author}{\bibfnamefont{H.}~\bibnamefont{Ryu}}, \bibinfo{author}{\bibfnamefont{S.}~\bibnamefont{Tang}}, \bibinfo{author}{\bibfnamefont{S.-K.} \bibnamefont{Mo}}, \bibnamefont{et~al.}, \bibinfo{journal}{Nature} \textbf{\bibinfo{volume}{564}}, \bibinfo{pages}{390} (\bibinfo{year}{2018}), ISSN \bibinfo{issn}{1476-4687}, \urlprefix\url{https://link.aps.org/doi/10.1038/s41586-018-0788-5}.

\bibitem[{\citenamefont{Dick et~al.}(2010)\citenamefont{Dick, Thelander, Samuelson, and Caroff}}]{10.1021/nl101632a}
\bibinfo{author}{\bibfnamefont{K.~A.} \bibnamefont{Dick}}, \bibinfo{author}{\bibfnamefont{C.}~\bibnamefont{Thelander}}, \bibinfo{author}{\bibfnamefont{L.}~\bibnamefont{Samuelson}}, \bibnamefont{and} \bibinfo{author}{\bibfnamefont{P.}~\bibnamefont{Caroff}}, \bibinfo{journal}{Nano Letters} \textbf{\bibinfo{volume}{10}}, \bibinfo{pages}{3494} (\bibinfo{year}{2010}), \urlprefix\url{https://doi.org/10.1021/nl101632a}.

\bibitem[{\citenamefont{Hjort et~al.}(2014)\citenamefont{Hjort, Lehmann, Knutsson, Zakharov, Du, Sakong, Timm, Nylund, Lundgren, Kratzer et~al.}}]{10.1021/nn504795v}
\bibinfo{author}{\bibfnamefont{M.}~\bibnamefont{Hjort}}, \bibinfo{author}{\bibfnamefont{S.}~\bibnamefont{Lehmann}}, \bibinfo{author}{\bibfnamefont{J.}~\bibnamefont{Knutsson}}, \bibinfo{author}{\bibfnamefont{A.~A.} \bibnamefont{Zakharov}}, \bibinfo{author}{\bibfnamefont{Y.~A.} \bibnamefont{Du}}, \bibinfo{author}{\bibfnamefont{S.}~\bibnamefont{Sakong}}, \bibinfo{author}{\bibfnamefont{R.}~\bibnamefont{Timm}}, \bibinfo{author}{\bibfnamefont{G.}~\bibnamefont{Nylund}}, \bibinfo{author}{\bibfnamefont{E.}~\bibnamefont{Lundgren}}, \bibinfo{author}{\bibfnamefont{P.}~\bibnamefont{Kratzer}}, \bibnamefont{et~al.}, \bibinfo{journal}{ACS Nano} \textbf{\bibinfo{volume}{8}}, \bibinfo{pages}{12346} (\bibinfo{year}{2014}), \urlprefix\url{https://doi.org/10.1021/nn504795v}.

\bibitem[{\citenamefont{Staudinger et~al.}(2020)\citenamefont{Staudinger, Moselund, and Schmid}}]{nanolett.9b04507}
\bibinfo{author}{\bibfnamefont{P.}~\bibnamefont{Staudinger}}, \bibinfo{author}{\bibfnamefont{K.~E.} \bibnamefont{Moselund}}, \bibnamefont{and} \bibinfo{author}{\bibfnamefont{H.}~\bibnamefont{Schmid}}, \bibinfo{journal}{Nano Letters} \textbf{\bibinfo{volume}{20}}, \bibinfo{pages}{686} (\bibinfo{year}{2020}), \urlprefix\url{https://doi.org/10.1021/acs.nanolett.9b04507}.

\bibitem[{\citenamefont{Dayeh et~al.}(2009)\citenamefont{Dayeh, Susac, Kavanagh, Yu, and Wang}}]{dayehStructural2009}
\bibinfo{author}{\bibfnamefont{S.~A.} \bibnamefont{Dayeh}}, \bibinfo{author}{\bibfnamefont{D.}~\bibnamefont{Susac}}, \bibinfo{author}{\bibfnamefont{K.~L.} \bibnamefont{Kavanagh}}, \bibinfo{author}{\bibfnamefont{E.~T.} \bibnamefont{Yu}}, \bibnamefont{and} \bibinfo{author}{\bibfnamefont{D.}~\bibnamefont{Wang}}, \bibinfo{journal}{Adv. Funct. Mater.} \textbf{\bibinfo{volume}{19}}, \bibinfo{pages}{2102} (\bibinfo{year}{2009}), \urlprefix\url{https://doi.org/10.1002/adfm.200801307}.

\bibitem[{\citenamefont{Li et~al.}(2014)\citenamefont{Li, Gan, McCartney, Liang, Yu, Yin, Yan, Gao, Wang, and Smith}}]{adma.201304021}
\bibinfo{author}{\bibfnamefont{L.}~\bibnamefont{Li}}, \bibinfo{author}{\bibfnamefont{Z.}~\bibnamefont{Gan}}, \bibinfo{author}{\bibfnamefont{M.~R.} \bibnamefont{McCartney}}, \bibinfo{author}{\bibfnamefont{H.}~\bibnamefont{Liang}}, \bibinfo{author}{\bibfnamefont{H.}~\bibnamefont{Yu}}, \bibinfo{author}{\bibfnamefont{W.-J.} \bibnamefont{Yin}}, \bibinfo{author}{\bibfnamefont{Y.}~\bibnamefont{Yan}}, \bibinfo{author}{\bibfnamefont{Y.}~\bibnamefont{Gao}}, \bibinfo{author}{\bibfnamefont{J.}~\bibnamefont{Wang}}, \bibnamefont{and} \bibinfo{author}{\bibfnamefont{D.~J.} \bibnamefont{Smith}}, \bibinfo{journal}{Advanced Materials} \textbf{\bibinfo{volume}{26}}, \bibinfo{pages}{1052} (\bibinfo{year}{2014}), \urlprefix\url{https://onlinelibrary.wiley.com/doi/abs/10.1002/adma.201304021}.

\bibitem[{\citenamefont{Becker et~al.}(2018)\citenamefont{Becker, Mork\"otter, Treu, Sonner, Speckbacher, D\"oblinger, Abstreiter, Finley, and Koblm\"uller}}]{PhysRevB.97.115306}
\bibinfo{author}{\bibfnamefont{J.}~\bibnamefont{Becker}}, \bibinfo{author}{\bibfnamefont{S.}~\bibnamefont{Mork\"otter}}, \bibinfo{author}{\bibfnamefont{J.}~\bibnamefont{Treu}}, \bibinfo{author}{\bibfnamefont{M.}~\bibnamefont{Sonner}}, \bibinfo{author}{\bibfnamefont{M.}~\bibnamefont{Speckbacher}}, \bibinfo{author}{\bibfnamefont{M.}~\bibnamefont{D\"oblinger}}, \bibinfo{author}{\bibfnamefont{G.}~\bibnamefont{Abstreiter}}, \bibinfo{author}{\bibfnamefont{J.~J.} \bibnamefont{Finley}}, \bibnamefont{and} \bibinfo{author}{\bibfnamefont{G.}~\bibnamefont{Koblm\"uller}}, \bibinfo{journal}{Phys. Rev. B} \textbf{\bibinfo{volume}{97}}, \bibinfo{pages}{115306} (\bibinfo{year}{2018}), \urlprefix\url{https://link.aps.org/doi/10.1103/PhysRevB.97.115306}.

\bibitem[{\citenamefont{Miao et~al.}(2012)\citenamefont{Miao, Yan, Van~de Walle, Lou, Li, and Chang}}]{PhysRevLett.109.186803}
\bibinfo{author}{\bibfnamefont{M.~S.} \bibnamefont{Miao}}, \bibinfo{author}{\bibfnamefont{Q.}~\bibnamefont{Yan}}, \bibinfo{author}{\bibfnamefont{C.~G.} \bibnamefont{Van~de Walle}}, \bibinfo{author}{\bibfnamefont{W.~K.} \bibnamefont{Lou}}, \bibinfo{author}{\bibfnamefont{L.~L.} \bibnamefont{Li}}, \bibnamefont{and} \bibinfo{author}{\bibfnamefont{K.}~\bibnamefont{Chang}}, \bibinfo{journal}{Phys. Rev. Lett.} \textbf{\bibinfo{volume}{109}}, \bibinfo{pages}{186803} (\bibinfo{year}{2012}), \urlprefix\url{https://link.aps.org/doi/10.1103/PhysRevLett.109.186803}.

\bibitem[{\citenamefont{Zhang et~al.}(2013)\citenamefont{Zhang, Lou, Miao, Zhang, and Chang}}]{PhysRevLett.111.156402}
\bibinfo{author}{\bibfnamefont{D.}~\bibnamefont{Zhang}}, \bibinfo{author}{\bibfnamefont{W.}~\bibnamefont{Lou}}, \bibinfo{author}{\bibfnamefont{M.}~\bibnamefont{Miao}}, \bibinfo{author}{\bibfnamefont{S.-c.} \bibnamefont{Zhang}}, \bibnamefont{and} \bibinfo{author}{\bibfnamefont{K.}~\bibnamefont{Chang}}, \bibinfo{journal}{Phys. Rev. Lett.} \textbf{\bibinfo{volume}{111}}, \bibinfo{pages}{156402} (\bibinfo{year}{2013}), \urlprefix\url{https://link.aps.org/doi/10.1103/PhysRevLett.111.156402}.

\bibitem[{\citenamefont{Zhang et~al.}(2014)\citenamefont{Zhang, Xu, Wang, Chang, and Zhang}}]{PhysRevLett.112.216803}
\bibinfo{author}{\bibfnamefont{H.}~\bibnamefont{Zhang}}, \bibinfo{author}{\bibfnamefont{Y.}~\bibnamefont{Xu}}, \bibinfo{author}{\bibfnamefont{J.}~\bibnamefont{Wang}}, \bibinfo{author}{\bibfnamefont{K.}~\bibnamefont{Chang}}, \bibnamefont{and} \bibinfo{author}{\bibfnamefont{S.-C.} \bibnamefont{Zhang}}, \bibinfo{journal}{Phys. Rev. Lett.} \textbf{\bibinfo{volume}{112}}, \bibinfo{pages}{216803} (\bibinfo{year}{2014}), \urlprefix\url{https://link.aps.org/doi/10.1103/PhysRevLett.112.216803}.

\bibitem[{\citenamefont{{See Supplemental Materials for the details of DFT calculations, the $k\cdot p$ Hamiltonian and the calculation of conductance and current with Refs.~\cite{KRESSE199615,PhysRevLett.77.3865,PhysRevB.59.1758,adma.201304021,PhysRevB.78.125116,Mao_2022,dayehStructural2009,10.1021/nn504795v,Zhang_2023, GAO2021107760,datta,kwant}}}()}]{SupplementalMaterials}
\bibinfo{author}{\bibnamefont{{See Supplemental Materials for the details of DFT calculations, the $k\cdot p$ Hamiltonian and the calculation of conductance and current with Refs.~\cite{KRESSE199615,PhysRevLett.77.3865,PhysRevB.59.1758,adma.201304021,PhysRevB.78.125116,Mao_2022,dayehStructural2009,10.1021/nn504795v,Zhang_2023, GAO2021107760,datta,kwant}}}}.

\bibitem[{\citenamefont{Zhang et~al.}(2023)\citenamefont{Zhang, Sheng, Song, Liang, Jiang, Sun, Wu, Weng, Fang, Dai et~al.}}]{Zhang_2023}
\bibinfo{author}{\bibfnamefont{S.}~\bibnamefont{Zhang}}, \bibinfo{author}{\bibfnamefont{H.}~\bibnamefont{Sheng}}, \bibinfo{author}{\bibfnamefont{Z.-D.} \bibnamefont{Song}}, \bibinfo{author}{\bibfnamefont{C.}~\bibnamefont{Liang}}, \bibinfo{author}{\bibfnamefont{Y.}~\bibnamefont{Jiang}}, \bibinfo{author}{\bibfnamefont{S.}~\bibnamefont{Sun}}, \bibinfo{author}{\bibfnamefont{Q.}~\bibnamefont{Wu}}, \bibinfo{author}{\bibfnamefont{H.}~\bibnamefont{Weng}}, \bibinfo{author}{\bibfnamefont{Z.}~\bibnamefont{Fang}}, \bibinfo{author}{\bibfnamefont{X.}~\bibnamefont{Dai}}, \bibnamefont{et~al.}, \bibinfo{journal}{Chinese Physics Letters} \textbf{\bibinfo{volume}{40}}, \bibinfo{pages}{127101} (\bibinfo{year}{2023}), \urlprefix\url{https://dx.doi.org/10.1088/0256-307X/40/12/127101,http://www.topmat.org/}.

\bibitem[{\citenamefont{Gao et~al.}(2021)\citenamefont{Gao, Wu, Persson, and Wang}}]{GAO2021107760}
\bibinfo{author}{\bibfnamefont{J.}~\bibnamefont{Gao}}, \bibinfo{author}{\bibfnamefont{Q.}~\bibnamefont{Wu}}, \bibinfo{author}{\bibfnamefont{C.}~\bibnamefont{Persson}}, \bibnamefont{and} \bibinfo{author}{\bibfnamefont{Z.}~\bibnamefont{Wang}}, \bibinfo{journal}{Computer Physics Communications} \textbf{\bibinfo{volume}{261}}, \bibinfo{pages}{107760} (\bibinfo{year}{2021}), ISSN \bibinfo{issn}{0010-4655}, \urlprefix\url{https://www.sciencedirect.com/science/article/pii/S0010465520303805}.

\bibitem[{\citenamefont{van Wees et~al.}(1988)\citenamefont{van Wees, van Houten, Beenakker, Williamson, Kouwenhoven, van~der Marel, and Foxon}}]{PhysRevLett.60.848}
\bibinfo{author}{\bibfnamefont{B.~J.} \bibnamefont{van Wees}}, \bibinfo{author}{\bibfnamefont{H.}~\bibnamefont{van Houten}}, \bibinfo{author}{\bibfnamefont{C.~W.~J.} \bibnamefont{Beenakker}}, \bibinfo{author}{\bibfnamefont{J.~G.} \bibnamefont{Williamson}}, \bibinfo{author}{\bibfnamefont{L.~P.} \bibnamefont{Kouwenhoven}}, \bibinfo{author}{\bibfnamefont{D.}~\bibnamefont{van~der Marel}}, \bibnamefont{and} \bibinfo{author}{\bibfnamefont{C.~T.} \bibnamefont{Foxon}}, \bibinfo{journal}{Phys. Rev. Lett.} \textbf{\bibinfo{volume}{60}}, \bibinfo{pages}{848} (\bibinfo{year}{1988}), \urlprefix\url{https://link.aps.org/doi/10.1103/PhysRevLett.60.848}.

\bibitem[{\citenamefont{Zhang et~al.}(2011)\citenamefont{Zhang, Cheng, Zhai, and Chang}}]{PhysRevB.83.081402}
\bibinfo{author}{\bibfnamefont{L.~B.} \bibnamefont{Zhang}}, \bibinfo{author}{\bibfnamefont{F.}~\bibnamefont{Cheng}}, \bibinfo{author}{\bibfnamefont{F.}~\bibnamefont{Zhai}}, \bibnamefont{and} \bibinfo{author}{\bibfnamefont{K.}~\bibnamefont{Chang}}, \bibinfo{journal}{Phys. Rev. B} \textbf{\bibinfo{volume}{83}}, \bibinfo{pages}{081402} (\bibinfo{year}{2011}), \urlprefix\url{https://link.aps.org/doi/10.1103/PhysRevB.83.081402}.

\bibitem[{\citenamefont{Wu et~al.}(2017)\citenamefont{Wu, Lin, Yang, Zhang, Shen, Lou, Yin, and Chang}}]{wuSpinpolarizedChargeTrapping2017a}
\bibinfo{author}{\bibfnamefont{Z.}~\bibnamefont{Wu}}, \bibinfo{author}{\bibfnamefont{L.}~\bibnamefont{Lin}}, \bibinfo{author}{\bibfnamefont{W.}~\bibnamefont{Yang}}, \bibinfo{author}{\bibfnamefont{D.}~\bibnamefont{Zhang}}, \bibinfo{author}{\bibfnamefont{C.}~\bibnamefont{Shen}}, \bibinfo{author}{\bibfnamefont{W.}~\bibnamefont{Lou}}, \bibinfo{author}{\bibfnamefont{H.}~\bibnamefont{Yin}}, \bibnamefont{and} \bibinfo{author}{\bibfnamefont{K.}~\bibnamefont{Chang}}, \bibinfo{journal}{RSC Adv.} \textbf{\bibinfo{volume}{7}}, \bibinfo{pages}{30963} (\bibinfo{year}{2017}), \urlprefix\url{https://doi.org/10.1039/C7RA03482B}.

\bibitem[{\citenamefont{Wakerly}(2008)}]{wakerly2008digital}
\bibinfo{author}{\bibfnamefont{J.~F.} \bibnamefont{Wakerly}}, \emph{\bibinfo{title}{Digital Design: Principles and Practices}} (\bibinfo{publisher}{Pearson Education}, \bibinfo{year}{2008}).

\bibitem[{\citenamefont{{The NI is also described by the same $k\cdot p$ Hamiltonian but with a different parameter $M=20$ meV $>0$ to obtain a NI phase in this ribbon.}}()}]{NI_para}
\bibinfo{author}{\bibnamefont{{The NI is also described by the same $k\cdot p$ Hamiltonian but with a different parameter $M=20$ meV $>0$ to obtain a NI phase in this ribbon.}}}

\bibitem[{\citenamefont{Datta}(2005)}]{datta}
\bibinfo{author}{\bibfnamefont{S.}~\bibnamefont{Datta}}, \emph{\bibinfo{title}{Quantum Transport: Atom to Transistor}} (\bibinfo{publisher}{Cambridge university press, Cambridge}, \bibinfo{year}{2005}).

\bibitem[{\citenamefont{Kim et~al.}(2020)\citenamefont{Kim, Myeong, Shin, Lim, Kim, Jin, Chang, Watanabe, Taniguchi, and Cho}}]{kim2020thickness}
\bibinfo{author}{\bibfnamefont{S.}~\bibnamefont{Kim}}, \bibinfo{author}{\bibfnamefont{G.}~\bibnamefont{Myeong}}, \bibinfo{author}{\bibfnamefont{W.}~\bibnamefont{Shin}}, \bibinfo{author}{\bibfnamefont{H.}~\bibnamefont{Lim}}, \bibinfo{author}{\bibfnamefont{B.}~\bibnamefont{Kim}}, \bibinfo{author}{\bibfnamefont{T.}~\bibnamefont{Jin}}, \bibinfo{author}{\bibfnamefont{S.}~\bibnamefont{Chang}}, \bibinfo{author}{\bibfnamefont{K.}~\bibnamefont{Watanabe}}, \bibinfo{author}{\bibfnamefont{T.}~\bibnamefont{Taniguchi}}, \bibnamefont{and} \bibinfo{author}{\bibfnamefont{S.}~\bibnamefont{Cho}}, \bibinfo{journal}{Nature nanotechnology} \textbf{\bibinfo{volume}{15}}, \bibinfo{pages}{203} (\bibinfo{year}{2020}), \urlprefix\url{https://doi.org/10.1038/s41565-019-0623-7}.

\bibitem[{\citenamefont{Kamaei et~al.}(2023)\citenamefont{Kamaei, Liu, Saeidi, Wei, Gastaldi, Brugger, and Ionescu}}]{kamaei2023ferroelectric}
\bibinfo{author}{\bibfnamefont{S.}~\bibnamefont{Kamaei}}, \bibinfo{author}{\bibfnamefont{X.}~\bibnamefont{Liu}}, \bibinfo{author}{\bibfnamefont{A.}~\bibnamefont{Saeidi}}, \bibinfo{author}{\bibfnamefont{Y.}~\bibnamefont{Wei}}, \bibinfo{author}{\bibfnamefont{C.}~\bibnamefont{Gastaldi}}, \bibinfo{author}{\bibfnamefont{J.}~\bibnamefont{Brugger}}, \bibnamefont{and} \bibinfo{author}{\bibfnamefont{A.~M.} \bibnamefont{Ionescu}}, \bibinfo{journal}{Nature Electronics} \textbf{\bibinfo{volume}{6}}, \bibinfo{pages}{658} (\bibinfo{year}{2023}), \urlprefix\url{https://doi.org/10.1038/s41928-023-01018-7}.

\bibitem[{\citenamefont{Liu et~al.}(2021)\citenamefont{Liu, Jaiswal, Shahi, Wei, Fu, Chang, Chakravarty, Liu, Yang, Liu et~al.}}]{acsnano.1c01503}
\bibinfo{author}{\bibfnamefont{M.}~\bibnamefont{Liu}}, \bibinfo{author}{\bibfnamefont{H.~N.} \bibnamefont{Jaiswal}}, \bibinfo{author}{\bibfnamefont{S.}~\bibnamefont{Shahi}}, \bibinfo{author}{\bibfnamefont{S.}~\bibnamefont{Wei}}, \bibinfo{author}{\bibfnamefont{Y.}~\bibnamefont{Fu}}, \bibinfo{author}{\bibfnamefont{C.}~\bibnamefont{Chang}}, \bibinfo{author}{\bibfnamefont{A.}~\bibnamefont{Chakravarty}}, \bibinfo{author}{\bibfnamefont{X.}~\bibnamefont{Liu}}, \bibinfo{author}{\bibfnamefont{C.}~\bibnamefont{Yang}}, \bibinfo{author}{\bibfnamefont{Y.}~\bibnamefont{Liu}}, \bibnamefont{et~al.}, \bibinfo{journal}{ACS Nano} \textbf{\bibinfo{volume}{15}}, \bibinfo{pages}{5762} (\bibinfo{year}{2021}), \urlprefix\url{https://doi.org/10.1021/acsnano.1c01503}.

\bibitem[{\citenamefont{Anandan et~al.}(2019)\citenamefont{Anandan, Nagarajan, Kakkerla, Yu, Ko, Singh, Lee, and Chang}}]{ANANDAN201930}
\bibinfo{author}{\bibfnamefont{D.}~\bibnamefont{Anandan}}, \bibinfo{author}{\bibfnamefont{V.}~\bibnamefont{Nagarajan}}, \bibinfo{author}{\bibfnamefont{R.~K.} \bibnamefont{Kakkerla}}, \bibinfo{author}{\bibfnamefont{H.~W.} \bibnamefont{Yu}}, \bibinfo{author}{\bibfnamefont{H.~L.} \bibnamefont{Ko}}, \bibinfo{author}{\bibfnamefont{S.~K.} \bibnamefont{Singh}}, \bibinfo{author}{\bibfnamefont{C.~T.} \bibnamefont{Lee}}, \bibnamefont{and} \bibinfo{author}{\bibfnamefont{E.~Y.} \bibnamefont{Chang}}, \bibinfo{journal}{Journal of Crystal Growth} \textbf{\bibinfo{volume}{522}}, \bibinfo{pages}{30} (\bibinfo{year}{2019}), \urlprefix\url{https://www.sciencedirect.com/science/article/pii/S0022024819303367}.

\bibitem[{\citenamefont{Seidl et~al.}(2021)\citenamefont{Seidl, Gluschke, Yuan, Tan, Jagadish, Caroff, and Micolich}}]{10.1021/acsnano.1c00483}
\bibinfo{author}{\bibfnamefont{J.}~\bibnamefont{Seidl}}, \bibinfo{author}{\bibfnamefont{J.~G.} \bibnamefont{Gluschke}}, \bibinfo{author}{\bibfnamefont{X.}~\bibnamefont{Yuan}}, \bibinfo{author}{\bibfnamefont{H.~H.} \bibnamefont{Tan}}, \bibinfo{author}{\bibfnamefont{C.}~\bibnamefont{Jagadish}}, \bibinfo{author}{\bibfnamefont{P.}~\bibnamefont{Caroff}}, \bibnamefont{and} \bibinfo{author}{\bibfnamefont{A.~P.} \bibnamefont{Micolich}}, \bibinfo{journal}{ACS Nano} \textbf{\bibinfo{volume}{15}}, \bibinfo{pages}{7226} (\bibinfo{year}{2021}), \urlprefix\url{https://doi.org/10.1021/acsnano.1c00483}.

\bibitem[{\citenamefont{Pirie et~al.}(2022)\citenamefont{Pirie, Sadhuka, Wang, Andrei, and Hoffman}}]{pirie_topological_2022}
\bibinfo{author}{\bibfnamefont{H.}~\bibnamefont{Pirie}}, \bibinfo{author}{\bibfnamefont{S.}~\bibnamefont{Sadhuka}}, \bibinfo{author}{\bibfnamefont{J.}~\bibnamefont{Wang}}, \bibinfo{author}{\bibfnamefont{R.}~\bibnamefont{Andrei}}, \bibnamefont{and} \bibinfo{author}{\bibfnamefont{J.~E.} \bibnamefont{Hoffman}}, \bibinfo{journal}{Phys. Rev. Lett.} \textbf{\bibinfo{volume}{128}}, \bibinfo{pages}{015501} (\bibinfo{year}{2022}), \urlprefix\url{https://link.aps.org/doi/10.1103/PhysRevLett.128.015501}.

\bibitem[{\citenamefont{Zhang et~al.}(2022)\citenamefont{Zhang, He, Zhang, Kong, Xu, and Zhang}}]{zhang_experimental_2023}
\bibinfo{author}{\bibfnamefont{F.}~\bibnamefont{Zhang}}, \bibinfo{author}{\bibfnamefont{L.}~\bibnamefont{He}}, \bibinfo{author}{\bibfnamefont{H.}~\bibnamefont{Zhang}}, \bibinfo{author}{\bibfnamefont{L.}~\bibnamefont{Kong}}, \bibinfo{author}{\bibfnamefont{X.}~\bibnamefont{Xu}}, \bibnamefont{and} \bibinfo{author}{\bibfnamefont{X.}~\bibnamefont{Zhang}}, \bibinfo{journal}{Laser \& Photonics Reviews} \textbf{\bibinfo{volume}{17}}, \bibinfo{pages}{2200329} (\bibinfo{year}{2022}), \urlprefix\url{https://onlinelibrary.wiley.com/doi/10.1002/lpor.202200329}.

\bibitem[{\citenamefont{Kresse and Furthmüller}(1996)}]{KRESSE199615}
\bibinfo{author}{\bibfnamefont{G.}~\bibnamefont{Kresse}} \bibnamefont{and} \bibinfo{author}{\bibfnamefont{J.}~\bibnamefont{Furthmüller}}, \bibinfo{journal}{Computational Materials Science} \textbf{\bibinfo{volume}{6}}, \bibinfo{pages}{15} (\bibinfo{year}{1996}), ISSN \bibinfo{issn}{0927-0256}, \urlprefix\url{https://www.sciencedirect.com/science/article/pii/0927025696000080}.

\bibitem[{\citenamefont{Perdew et~al.}(1996)\citenamefont{Perdew, Burke, and Ernzerhof}}]{PhysRevLett.77.3865}
\bibinfo{author}{\bibfnamefont{J.~P.} \bibnamefont{Perdew}}, \bibinfo{author}{\bibfnamefont{K.}~\bibnamefont{Burke}}, \bibnamefont{and} \bibinfo{author}{\bibfnamefont{M.}~\bibnamefont{Ernzerhof}}, \bibinfo{journal}{Phys. Rev. Lett.} \textbf{\bibinfo{volume}{77}}, \bibinfo{pages}{3865} (\bibinfo{year}{1996}), \urlprefix\url{https://link.aps.org/doi/10.1103/PhysRevLett.77.3865}.

\bibitem[{\citenamefont{Kresse and Joubert}(1999)}]{PhysRevB.59.1758}
\bibinfo{author}{\bibfnamefont{G.}~\bibnamefont{Kresse}} \bibnamefont{and} \bibinfo{author}{\bibfnamefont{D.}~\bibnamefont{Joubert}}, \bibinfo{journal}{Phys. Rev. B} \textbf{\bibinfo{volume}{59}}, \bibinfo{pages}{1758} (\bibinfo{year}{1999}), \urlprefix\url{https://link.aps.org/doi/10.1103/PhysRevB.59.1758}.

\bibitem[{\citenamefont{Ferreira et~al.}(2008)\citenamefont{Ferreira, Marques, and Teles}}]{PhysRevB.78.125116}
\bibinfo{author}{\bibfnamefont{L.~G.} \bibnamefont{Ferreira}}, \bibinfo{author}{\bibfnamefont{M.}~\bibnamefont{Marques}}, \bibnamefont{and} \bibinfo{author}{\bibfnamefont{L.~K.} \bibnamefont{Teles}}, \bibinfo{journal}{Phys. Rev. B} \textbf{\bibinfo{volume}{78}}, \bibinfo{pages}{125116} (\bibinfo{year}{2008}), \urlprefix\url{https://link.aps.org/doi/10.1103/PhysRevB.78.125116}.

\bibitem[{\citenamefont{Mao et~al.}(2022)\citenamefont{Mao, Yan, Xue, Ai, Yang, Cui, Yuan, Ren, and Miao}}]{Mao_2022}
\bibinfo{author}{\bibfnamefont{G.-Q.} \bibnamefont{Mao}}, \bibinfo{author}{\bibfnamefont{Z.-Y.} \bibnamefont{Yan}}, \bibinfo{author}{\bibfnamefont{K.-H.} \bibnamefont{Xue}}, \bibinfo{author}{\bibfnamefont{Z.}~\bibnamefont{Ai}}, \bibinfo{author}{\bibfnamefont{S.}~\bibnamefont{Yang}}, \bibinfo{author}{\bibfnamefont{H.}~\bibnamefont{Cui}}, \bibinfo{author}{\bibfnamefont{J.-H.} \bibnamefont{Yuan}}, \bibinfo{author}{\bibfnamefont{T.-L.} \bibnamefont{Ren}}, \bibnamefont{and} \bibinfo{author}{\bibfnamefont{X.}~\bibnamefont{Miao}}, \bibinfo{journal}{Journal of Physics: Condensed Matter} \textbf{\bibinfo{volume}{34}}, \bibinfo{pages}{403001} (\bibinfo{year}{2022}), \urlprefix\url{https://dx.doi.org/10.1088/1361-648X/ac829d}.

\bibitem[{\citenamefont{Groth et~al.}(2014)\citenamefont{Groth, Wimmer, Akhmerov, and Waintal}}]{kwant}
\bibinfo{author}{\bibfnamefont{C.~W.} \bibnamefont{Groth}}, \bibinfo{author}{\bibfnamefont{M.}~\bibnamefont{Wimmer}}, \bibinfo{author}{\bibfnamefont{A.~R.} \bibnamefont{Akhmerov}}, \bibnamefont{and} \bibinfo{author}{\bibfnamefont{X.}~\bibnamefont{Waintal}}, \bibinfo{journal}{New Journal of Physics} \textbf{\bibinfo{volume}{16}}, \bibinfo{pages}{063065} (\bibinfo{year}{2014}), \urlprefix\url{https://dx.doi.org/10.1088/1367-2630/16/6/063065}.

\end{thebibliography}


\end{document}